\documentclass[12pt]{article}
\usepackage{graphicx,color}
\usepackage[labelformat=empty]{caption}
\newcommand\zed{{-\zeta}}

\begin{document}

\title{A minimal physical model captures the shapes of crawling cells}

\author{E. Tjhung$\dagger$, A. Tiribocchi$\dagger$, D. Marenduzzo and M. E. Cates \\ {\small $\dagger$: equal contributions} \\
SUPA, School of Physics and Astronomy,\\ University of Edinburgh, JCMB Kings Buildings,\\ Mayfield Road, Edinburgh EH9 3JZ, United Kingdom}
\date{}

\maketitle

{\bf Cell motility in higher organisms (eukaryotes) is crucial to biological functions ranging from wound healing to immune response, and also implicated in diseases such as cancer. For cells crawling on hard surfaces, significant insights into motility have been gained from experiments replicating such motion in vitro. Such experiments show that crawling uses a combination of actin treadmilling (polymerization), which pushes the front of a cell forward, and myosin-induced stress (contractility), which retracts the rear. Here we present a simplified physical model of a crawling cell, consisting of a droplet of active polar fluid with contractility throughout, but treadmilling connected to a thin layer near the supporting wall. The model shows a variety of shapes and/or motility regimes, some closely resembling cases seen experimentally. Our work strongly supports the view that cellular motility exploits autonomous physical mechanisms whose operation does not need continuous regulatory effort.
}

How do living cells move around within their environment, whether {\it in
vitro} (e.g., on a glass slide), or {\it in vivo} (e.g., within a tissue)? Is their motion sustained by continuous action of the cell's complex biochemical networks, or do these issue instructions to a physically autonomous engine of motility?
Experiments that replicate cellular motility with simplified {\it in vitro} systems have gone some way towards answering these questions~\cite{alberts,bray,Bausch,Carlier}.

Cellular self-propulsion in 3D gels or tissues might involve either `grabbing' of the extracellular matrix~\cite{bray}, or `swimming' via 
actomyosin flow in the cell interior~\cite{poincloux,rhoda,elsen}.
But, since 3D motility is hard to study experimentally, detailed mechanisms remain largely speculative. Much better understood is a case we address here: cells moving on a 2D solid substrate. 

The crawling motion of such cells combines two processes~\cite{bray}. One is actin polymerization, in which monomers add to one end of an actin filament and leave from the other. Given a net polarization of the actin network, this `treadmilling' pushes the leading edge of the cell forwards. The second process is the `walking' of myosin motors along filaments. This creates a contractile~\cite{sriram} stress that, alongside interfacial tension and/or membrane elasticity, opposes cellular elongation. Contractility thus assists retraction of the rear part of a crawling cell (which includes the nucleus). 
Also implicated in crawling are assemblies of intracellular membrane proteins called focal adhesions~\cite{alberts}, which enhance mechanical connectivity between actomoyosin and the solid substrate. These increase the effective wall friction and improve crawling efficiency (just as a child can crawl more easily on a pile carpet than a polished floor). Focal adhesions also signal alterations to the balance between treadmilling and contractility~\cite{demali,cooper}.
Notably though, significant motility is maintained even in laser-ablated or reconstituted cell fragments that lack most or all of the cell's regulatory machinery~\cite{fragments}.

Our goal here is to set up a minimal model of crawling motility, to see how fully the observed behaviour can be understood using physics alone. 
To achieve this goal, we intentionally rely on a very simplified description of complex biochemical and biophysical processes such as F-actin polymerisation and actomyosin contractility. 
We also employ only an elementary description of the cell membrane, as well as of its coupling to the interior and exterior of the cell.
The advantages of such an approach is that there are arguably as few free parameters as possible within the model, and this allows us to isolate the key physical elements of cell crawling (and/or spreading~\cite{sriram2}) at surfaces. In other words, we can directly answer the question posed above: can cellular motility be seen as a physics-based machinery which is coupled to, and controlled by, a complicated underlying biochemical network, but is nevertheless able to function independently? The alternative is that the biochemical detail is fundamental, and no aspects of it can be disregarded, even for a basic understanding of motility. This is an important question, both to understand the basic mechanism of cell motility, and to inform subsequent biomimetic work intending to replicate cellular behaviour with synthetic components. In this respect, our simulation model can be viewed as the {\it in silico} analogue of {\it in vitro} experiments with cell extracts and cell fragments~\cite{fragments}.

The drawback inherent within our approach is that, from the beginning, some of the important biological detail is lost. One can argue that, if experimental observations are reproduced by the minimal framework, this detail may be inessential to the fundamental biophysics of motility, and this by itself is very useful information. Nevertheless, in order to check the picture and to prove the previous statement unambiguously, one should eventually insert back realistic details one by one, and explicitly compare the results obtained with those of the fully simplified, minimal model. We initiate this last approach here as well, for a few cell biological features that our baseline model does not capture well. These include the allowance for a relatively high membrane viscosity, and for wall slip, or equivalently, retrograde flow~\cite{Vallott}; see Supplementary Note 1, 2, 3 for details of these checks.

In our minimal model, we represent a cell or cell extract as a droplet of active polar fluid (actomyosin) immersed in a second fluid and confined by interfacial tension~\cite{elsen}. 
(The latter is similar to, but simpler than, the fixed area constraint imposed by the outer membrane of a real cell.) As in real cells, actin treadmilling is confined both to the leading edge of the cell, and within a finite thickness $\lambda$ from the substrate.
Our droplet interacts with a solid wall through a controllable friction, represented by a partial-slip boundary condition. 
In contrast to most previous computer models for cell motility~\cite{elsen,ziebert,ziebert2,casademunt,callanjones,mogilner,mogilner2,levine1,levine2}, ours treats cells as 3D objects. Previous 2D models often incorporated a number of important biological details, such as variable adhesion with the substrate~\cite{ziebert2}, or the dynamics of signalling proteins~\cite{mogilner}. A difference with our framework, besides the general philosophy highlighted above, is that these 2D models typically view the cell from above and cannot resolve the flow within the cell, or the presence or absence of slip at the boundary, whereas our model can resolve these features. Importantly, our 3D framework also allows us to directly address the factors influencing the lamellipodium shape, which is the most commonly observed morphology in crawling cells. We are aware of one previous 3D study~\cite{herant} which recreated the motion of flat cell fragments. Ref.~\cite{herant} differs from our approach in its aims, as it focuses on a specific type of cell motility, rather than attempting to describe morphological transitions between non-motile and motile cells or between different kinds of motility, as we do here.

\section*{Results}
{\bf Crawling dynamics in quasi-2D.} To begin, we consider a quasi-2D geometry describing a thin slice through a crawling droplet (Fig.~1a). 
At the wall we assume parallel anchoring of actin polarization (represented by black arrows in the figure), and also a no-slip boundary condition on fluid flow as expected when focal adhesions are plentiful
(the effect of a small slip is considered below).
In these quasi-2D simulations actin polarization (hence treadmilling) is confined both to the leading edge of the cell, and within a finite thickness $\lambda$ from the substrate.
This is to model real cells where actin filaments are present, and polarized, mainly in a cortex towards the exterior~\cite{verkhovsky}. Importantly, no qualitative differences in the results are observed if we choose 
instead a simpler representation of treadmilling, where this is no longer localised at the leading edge, but still confined within a thickness $\lambda$ from the substrate, see Supplementary Fig.~1.
At the droplet's interface with the surrounding fluid we impose outward normal anchoring of variable strength $\beta$. 
The equations of motion are as previously published~\cite{elsen}, and detailed in the Methods Section.

Fig.~1a shows two snapshots of such a quasi-2D droplet in steady state. 
Upon increasing the treadmilling rate $w_0$, we observe progression from a 
droplet that moves along the wall with only modest distortion, 
to one that extends a thin protrusion at its leading edge (also see Supplementary 
Movies 1 and 2). 
In both regimes the droplet moves with a speed close to $w_0$. 
Fig.~1b plots the steady-state wall contact area against $w_0$, in the case of no-slip boundary conditions ($s=0$, red/plusses curve) and partial slip ($s=0.02$, $s=0.04$, $s=0.06$, green/crosses,  
blue/asterisks and purple/squares curves respectively).
The parameter $s$ controls the amount of slip of the fluid at the solid wall, and is defined such that $s=0$ corresponds to no-slip while $s=1$ corresponds to full-slip (see Supplementary Note 1
for its mathematical definition). 
The parameter $s$ is thus inversely related to the density of focal adhesions between the substrate and the cell.
In both cases, with or without wall slip, the contact area increases sharply near a critical value $w_c(s)$, suggestive of a dynamical transition or sharp crossover between two regimes.
This transition becomes sharper as the amount of slip on the substrate augments (green/crosses curve), and
discontinuous at critical $w_c$ for higher values of $s$ (blue/asterisks and purple/squares curves). The strong hysteresis of the last curve
confirms its discontinuous nature.
Note that in these quasi-2D simulations, we have set actomyosin contractility to zero. This suggests that actomyosin contraction is not essential to the crawling mechanism; indeed, including $\zeta$ leads to 
almost unchanged results (see Supplementary Fig.~1).
Hence the protruded shape observed in crawling cells can indeed be achieved by just a balance of actin treadmilling against interfacial tension at the cell membrane.
Finally in Fig.~1c, the crawling speed of our droplet (in steady state) is plotted against the slip parameter $s$ while the treadmilling rate is fixed.
As expected, increasing the amount of slip strongly reduces droplet motility via the crawling mechanism.
Additionally, we still observe a discontinuous transition at critical slip $s_c$ above which no protrusion is observed any longer.
This concurs with a previous comparison of crawling with contractility-driven `cell swimming', in which the wall friction was crudely represented by a depth-integrated drag term~\cite{elsen}.

{\bf Crawling dynamics in 3D.} The quasi-2D simulations in Fig.~1 give intriguing hints to the physics of lamellipodial protrusions, which are the pancake like structures often seen at the leading edge of crawling cells.
However, the structure of such cells is fully three-dimensional, and the lamellipodium is only one of several morphologies attained by live cells, whether crawling or stationary, when attached to a solid support.
We have therefore performed full 3D simulations of a slight variation of the same minimal model as we used in 2D. 
With respect to the 2D case, in the 3D case the effect of actomyosin contractility is much stronger, so we present results only with it switched on from the beginning.
Unlike actin treadmilling, actomyosin contractility is known to show highest activity deep inside the cell, close to the cell nucleus, rather than at the cell periphery~\cite{verkhovsky}.
To simplify the model in 3D, actin polarization (and thus contractility) is therefore now assumed to be uniform throughout the droplet. Additionally, actin polarization is no longer confined to the leading edge of the cell, 
but we retain a spatially varying treadmilling rate, $w_0\rightarrow w(z)=w_0\exp(-z/\lambda)$, to confine actin treadmilling close to the substrate.
The use of this simplified model, where the active droplet is homogeneously polarised, is motivated by our finding above that the localisation of actin polymerisation is not essential to observe the morphological transition between
non-motile and motile cells in Fig.~1 (see Supplementary Fig.~1). 

As we shall see, our minimal 3D model supports the formation of various dynamical cell shapes, some with surprisingly direct experimental counterparts.
Fig.~2 shows a selection of these steady state morphologies, together with a direct comparison with experimental morphologies taken from the literature. 

Fig.~2a-c shows a droplet subject to strong normal anchoring of the polarity (large $\beta$). 
This boundary condition, which might model the promoted nucleation of outwardly polarized actin by Arp2/3 complexes close to the outer membrane~\cite{bray}, 
enforces radial polarization with an `aster' defect at the centre~\cite{aster}. 
Treadmilling then causes the droplet to spread, but not to move. In steady state it adopts a symmetric `fried egg' 
whose shape resembles that of an inactive platelet~\cite{alberts} or a stationary 
keratocyte~\cite{Yam2007}. (We speculate that adding boundary pinning of the cell perimeter, and/or radial modulation of the treadmilling strength $w_0$, might create shapes more closely resembling mesenchymal stem cells~\cite{mesenchymal}). 

Figs.~2d-f, 2g-i, 2j-l show steady-state cell shapes for (rightward) crawling cells; the morphology is selected by a combination of treadmilling rate $w_0$, contractility $-\zeta$, and anchoring strength $\beta$. Fig.~2d-f (see also Supplementary Movie 3) arises when all three effects compete; this closely resembles a motile keratocyte (see e.g.~\cite{bray,Barnhart2011}). The pronounced lamellipodium at the leading edge stems from a balance between contractility-induced splay of the polarity field~\cite{elsen,sriram}, and its soft anchoring at the leading edge. The prediction of this familiar dynamical structure from such minimal physical ingredients is a major finding of this paper. 
(In contrast the overhang at the cell rear is a detail, and can be eliminated if actin polarization or treadmilling rate $w(x,y,z)$ is set to zero at the back of the cell, as done e.g. in Fig.~1a.)

The cell shape in Fig.~2g-i arises when the anchoring strength $\beta$ is decreased further: contractility-induced splay now dominates, creating an indented structure possibly reminiscent of the `phagocytic cup' which forms when a cell needs to engulf a solid particle~\cite{Dembo}. (Such cups are typically not planar but have been reported for crawling as well as free cells~\cite{cupcrawl,JCS1}.)
Finally, Fig.~2j-l shows a droplet where treadmilling is dominant: both the anchoring and contractility are much weaker than in 2d-f or 2g-i. Importantly, without those contributions the cell shape no longer shows the keratocyte-like lamellipodium morphology. Instead, the leading-edge forms a pointed protrusion, qualitatively resembling a `pseudopod', as often reported in ameoboid cell motion both on surfaces and in 3D geometries~\cite{leukocyte1,JCS2}. 

{\bf Phase diagram.} The results shown in Fig.~2 were obtained by varying  the treadmilling parameter $w_0$, the contractile activity $-\zeta$, and also the anchoring 
$\beta$ which controls the orientation of the polarisation at the cell boundary. 
Although a systematic study of the phase diagram in this multidimensional parameter space is extremely demanding computationally, and hence
outside the scope of the current work, we observe (see Fig.~3)
that similar morphologies can also be obtained by keeping the value of $w_0$ constant, varying $\beta$,
and adjusting $\zeta$ to be as large as possible avoiding at
the same time droplet breakup (which occurs when $\zeta$ is too large; 
see also Ref.~\cite{giomi} where a distinct but related study of
contractility-driven breakup 2D active droplets is presented). 

For very small values of $\beta$, a finger-like structure (the ``pseudopodium'' in Fig.~2) is dominant, while fan-shaped 
geometries (such as the ``lamellipodium'' in Fig.~2) emerge
when $\beta$ is larger, due to the higher splay 
distortion of the actin filaments arising at the leading edge of the cell. 
For very strong anchoring, the lamellipodium morphology becomes unstable,
and the cell attains a fried-egg shape, which is non-motile. 
Experiments (see e.g. Ref.~\cite{Barnhart2011}) have studied how
cell shape is affected by adhesion to the substrate. In our model an increase in the adhesion 
would affect both the treadmilling and the anchoring, decreasing the first and increasing the second. 
Interestingly, our simulations suggest that lamellipodia are found for intermediate values of the 
anchoring strength and are disfavoured when treadmilling decreases, 
qualitatively confirming what has been experimentally observed by 
Barnhart et al.~\cite{Barnhart2011}.

{\bf Evolution of polarisation and velocity fields.}
Besides the concentration field whose structure is shown in Fig.~2, our model also tracks the evolution of the polarization and intracellular flow fields, ${\mathbf P}$ and ${\mathbf u}$ respectively.

In Fig.~4 we show the polarization field close to the substrate (left column) and the flow field (right column) corresponding to the 3D steady state morphologies reported in  Fig.~2.
The polarization field shows a transition between the aster corresponding to the `fried-egg' state (Fig.~4a) and polarized uniform (Fig.~4d) and splayed (Fig.~4b,c) pattern;
these are qualitatively in broad agreement with the experimental data for keratocytes presented 
in~\cite{Yam2007}, where it was established that motility requires symmetry breaking of the actin fibre orientation
field, from an aster like conformation as in our ``fried-egg'' morphology (Fig.~4a)
to an asymmetric conformation in a lamellipodium, where the barbed ends of actin fibres are approximately normal 
to the cell membrane at the leading edge (as in our simulations in Fig.~4c).

Coming to the flow field, we first note that ${\mathbf u}$ (plotted in Fig.~4) should be thought of as an 
intracellular solvent flow inside the cell, which has not been measured often in experiments (an exception is
Ref.~\cite{keren} which is discussed in the Supplementary Note 2). This intracellular flow is quite distinct from the flow of F-actin, 
which is usually tracked in cell biology experiments, and which is commonly referred to as ``retrograde flow''~\cite{Vallott}.
F-actin flows inwards within the cell, from the leading edge to the bulk. 
The velocity field for the ``fried-egg'' cell is characterized by 
a quadrupolar velocity profile (in a fully 3D view), more intense
close to the surface of the droplet where the curvature is higher than near the substrate (see Fig.~4a).
The flow in the vicinity of the membrane is not unlike that measured in the case of an activity-driven
motile tumour cell ``swimming'' within a 3D matrigel~\cite{poincloux}, and even closer correspondence is found between 
that experiment and our contractility-driven droplets (which move in the absence of treadmilling, see Fig.~5).
The symmetric structure of the flow field in the ``fried-egg'' morphology explains its non-motile character.
The flow field for the keratocyte, lamellipodium-like shape, and for the phagocytic cup (Fig.~4b-c) is instead 
unidirectional (along the direction of motion) in the laboratory frame of reference, and gently follows the fan-out structure of the surface cell profile, 
being more intense close to both cytoplasmic projections. 
A similar structure is seen for the ``pseudopodium'' (Fig.~4d), with the difference that the flow is now almost 
entirely localised around the finger-like protrusion. Interestingly the highest magnitude of the fluid flow for the last three 
cases is twice larger than the one for the fried egg cell. 

The flow fields of these structures in are consistent with previous computer simulations~\cite{mogilnerflow}, and with the flow fields mapped experimentally in Ref.~\cite{keren} with decreased contractility compared to the wild-type (see also Supplementary Fig.~2).
Experiments suggest that contractility can propel intracellular flow significantly, and 2D simulations with variable $\zeta$ in our model confirm this qualitatively (see Supplementary Fig.~3 and Supplementary Note 2  for further discussion).

{\bf Oscillatory dynamics and medium-dependent viscosity.} Not all parameters in our model lead to steady state shapes; we also find spontaneous oscillations in shape and motility (see Supplementary Fig.~4, Supplementary Note 3, and Supplementary Movie 4). These or similar oscillations have long been reported experimentally~\cite{leukocyte3} and it is remarkable that they can emerge in principle from a  model that is devoid of all biochemical feedbacks.

All our simulations address the case where the active droplet is surrounded by a fluid of similar viscosity to its interior. This is an adequate model of wall-bound cells moving through an aqueous medium as is normally the case 
{\it in vitro} and {\it in vivo}, where cells are surrounded by at least a thin film of fluid. Accounting for the higher viscosity of the membrane also leads to similar results; we have performed selective numerical checks of this (Supplementary Fig.~5).

{\bf Crawling in the absence of treadmilling.} Motility of the crawling droplets in Fig.~2d-f, 2g-i, 2j-l is powered by
treadmilling, although contractility was found to play a crucial role in determining cell shape. What happens if treadmilling is absent altogether? It was shown in~\cite{elsen} that when suspended in bulk fluid a contractile droplet can `swim' by undergoing spontaneous splay. This creates an internal toroidal fluid flow which entrains the external fluid and leads to self-propulsion. Our simulations show that the same physical mechanism is still available for a fluid-immersed droplet attached to a wall (Fig.~4). The flow associated with
such contractility-driven active droplets is similar to that observed
experimentally in cells `swimming' in 3D matrigel (Fig.~5).

\section*{Discussion}
In conclusion, we have presented a minimal physical model of cell crawling, based on an active fluid droplet, attached to a solid wall and immersed in a second fluid. Key ingredients in the model are
actin treadmilling (confined to a thin layer of depth $\lambda$ close to the substrate), 
actomyosin contractility (throughout the droplet), full anchoring
of the actin polarity parallel to the wall, and its partial anchoring normal to the fluid-fluid interface. Hydrodynamic fluid flow is properly treated throughout.
We showed that on increasing the
treadmilling rate, such a droplet undergoes a nonequilibrium 
morphological transition to form a protrusion layer whose thickness is set by $\lambda$. Variation of the treadmilling, contractility and anchoring parameters lead in 3D to a striking series of cell morphologies, resembling
structures seen in real eukaryotic cells including both 
lamellipodia and pseudopods.
We also find regimes where oscillatory modulations of shape and motility arise.
Our results show that surprisingly many of the observed features of motile and spreading cells (though of course not all of them) can emerge from relatively simple physical ingredients.
This suggests that some aspects of cellular motility might fruitfully be thought of as driven by a physics-based `motility engine' whose function, although controlled by the cell's complex biochemical feedback networks, does not directly depend upon these for its basic operational principles. It would be of interest to extend the current model to study cell motility inside complex media, such as the extracellular matrix, or within tissues~\cite{jens}.

\section*{Methods}

We briefly outline here the hydrodynamic model used in this work.
We consider a droplet of cellular matter sitting on a solid substrate as a 
simple representation of a crawling cell. 
The droplet phase is represented by a scalar field $\phi(\mathbf{r},t)>0$.
Outside the droplet, we have a passive isotropic fluid with 
$\phi(\mathbf{r},t)=0$ to represent the wet environment around the cell and 
the substrate.
Secondly we define a polarization field $\mathbf{P}(\mathbf{r},t)$ to 
represent the actin filaments which are polarized and abundant mainly at the 
cortex of the cell. 
More specifically, $\mathbf{P}$ is the coarse grained average of all the 
orientations of the actin filaments: $\mathbf{P}=\left<\mathbf{p}\right>$
where $\mathbf{p}$ is a unit tangent oriented from the end at which the actin 
filaments depolymerize to the opposite end at which they polymerize.
Finally, we also have the average velocity field $\mathbf{v}(\mathbf{r},t)$ of both the cellular matter and the solvent.

\medskip
\noindent
\textbf{Underlying free energy.}
Following the phenomenological approach used in Refs.~\cite{elsen} and ~\cite{Ramaswamy},
we first introduce a free-energy functional that
controls the equilibrium physics of a polar liquid-crystalline passive droplet
\begin{eqnarray}\label{fe}
F[\phi,\mathbf{P}] 
&=&\int d^{3}r\,\{ \frac{a}{4\phi_{cr}^{4}}\phi^{2}(\phi-\phi_{0})^{2}+\frac{k}{2}\left|\nabla \phi\right|^{2} \nonumber\\
&-&\frac{\alpha}{2}  \frac{(\phi t(\mathbf{r})-\phi_{cr})}{\phi_{cr}}  \left|\mathbf{P}\right|^2+ \frac{\alpha}{4}\left|\mathbf{P}\right|^{4}+\frac{\kappa}{2}(\nabla\mathbf{P})^{2}\nonumber\\
&+&\beta\mathbf{P}\cdot\nabla\phi\}.
\end{eqnarray}
The first two terms of Eq.~(\ref{fe}) control bulk and interfacial properties of the droplet. 
In particular, $k>0$ determines the droplet interfacial tension and $a>0$ is a phenomenological constant. 
The bulk term is chosen to stabilise a droplet phase ($\phi\simeq\phi_0$) immersed in a second fluid ($\phi=0$) and $\phi_{cr}=\phi_0/2$. 
The parameter $t(\mathbf{r})$ controls a second order transition from isotropic ($|\mathbf{P}|=0$) to polar phase ($|\mathbf{P}|>0$) inside the droplet 
as the value of $t$ changes from $0$ to $1$.
Note the for all our 3D work, and for selected cases in 2D (see 
Supplementary Fig.~1),  
we have taken $t({\mathbf r})=1$ everywhere for simplicity; we discuss
some implications of this choice below.
The functional form of $t(\mathbf{r})$ for the remaining quasi-2D simulations is:
\begin{equation}
t(y,z) = \frac{1}{4} \left(1+\tanh(y-y_{back}-\Lambda)\right) \left(1-\tanh(z-\lambda)\right) 
\end{equation}  
where $z$ is the vertical distance from the substrate, $y$ is the coordinate parallel to the substrate and $y_{back}$ is the position of the cell rear.
Basically, the value of $t$ is equal to $1$ when $z$ is less than some thickness $\lambda$ from the substrate and at least a distance $\Lambda$ from the cell rear.
This defines the polarized (hence active) region at the leading edge of the droplet where actin treadmilling takes place
(see black arrows in Fig.~1A).
Incidentally, $\lambda$ also sets the lengthscale for the thickness of the protrusion layer.
Finally, elsewhere in the droplet $t$ is equal to $0$ which gives rise to an isotropic phase (and hence no actin treadmilling).

In summary, there are three equilibrium phases which can be obtained by minimizing the free energy $F[\phi,\mathbf{P}]$ in a state of uniform $\phi$ and $\mathbf{P}$, setting
$\frac{\delta F}{\delta \phi}=0$ and $\frac{\delta F}{\delta {\bf P}}=0$.
These three phases are: the isotropic passive fluid, external to the droplet ($\phi=0$ and $\mathbf{P}=0$);
the polarized region of the droplet, corresponding to the cortical actin-myosin network ($\phi=\phi_{eq1}$ and $\mathbf{P}=\mathbf{P}_{eq}$); and
the isotropic region of the droplet, corresponding to cell nucleus and other cellular materials which do not play a role in cell motility 
($\phi=\phi_{eq2}$ and $\mathbf{P}=0$).

The other three terms in the free energy describe bulk and elastic properties of the polar field, where $\alpha$ is a positive (for stability) constant and $\kappa$ is the elastic constant
(in the one-elastic constant approximation~\cite{PGG}) that describes the energetic cost due to elastic deformation in a liquid-crystalline environment. 
The surface tension $\gamma$ will depend on $a$, $k$ and on the elastic constant $\kappa$. Across the droplet interface, the values of the concentration and the polarization fields
vary smoothly from $\phi=\phi_{eq}$ and  ${\bf P}={\bf P}_{eq}$ to $\phi=0$ and ${\bf P}=0$. In general the width of the diffuse interface also depends on the same parameters. In our simulations,
the interfacial width is typically much smaller than the size of the droplet.
Finally the last term $\beta\mathbf{P}\cdot\nabla\phi$
takes into account the anchoring of $\mathbf{P}$ to the droplet surface (so that for $\beta>0$ the polar field $\mathbf{P}$ preferentially points perpendicularly outwards at the droplet perimeter
while for $\beta<0$ it preferentially points perpedicularly inwards).

\medskip
\noindent
\textbf{Equations of motion.}
The velocity field $\mathbf{v}(\mathbf{r},t)$ obeys the continuity and the Navier-Stokes equations, which in the incompressible limit are
\begin{eqnarray}
\nabla\cdot\mathbf{v} & = & 0\\
\rho\left(\frac{\partial}{\partial t}+\mathbf{v}\cdot\nabla\right)\mathbf{v} & = & -\nabla P+\nabla\cdot(\underline{\underline{\sigma}}^{active}+\underline{\underline{\sigma}}^{passive}).
\end{eqnarray}
Here $P$ is the isotropic pressure and $\underline{{\underline{\sigma}}}^{active}+\underline{{\underline{\sigma}}}^{passive}$ is the total stress, given by the sum of active and passive contributions. 
The first of them is
\begin{equation}
\sigma_{\alpha\beta}^{active}=\zed \phi P_{\alpha}P_{\beta}\label{eq:active-stress},
\end{equation}
that can be derived by summing the contributions from each force dipole and coarse graining~\cite{Ramaswamy}. The constant $\zeta$ is the activity parameter which is negative for
contractile particles and positive for extensile particles. Its magnitude $|\zeta|$ is proportional to the strength of the force dipole. 
For the case of actomyosin networks, $\zeta$ is negative signifying the tendency to contract along the direction of actin filaments.
The strength of actomyosin contraction is thus given by its magnitude $|\zeta|$.
As we have seen from Fig.~2, actomyosin contractility is responsible for the formation of a variety of complex 3D structures in crawling cells.
Also note that in our calculations above, 
we neglect the contributions of higher order terms (such as $\partial_{\alpha}P_{\beta}$) in the active stress, although these are in principle allowed by symmetry~\cite{Marc-Liver}.
This assumption is appropriate for actomyosin solutions, which are made up by elongated fibres of uniform width whose degree of asymmetry between the particle's head and tail is supposed to be small.

We note that actomyosin contraction shows highest 
activity at the cell rear~\cite{verkhovsky} rather than at the leading edge.
This is in contrast to actin treadmilling which shows highest activity at 
the leading edge of the cell.
Thus to study the global effect of actomyosin contractility in 3D, 
we make a further simplification by assuming 
actin polarization to be homogenous and uniform inside the droplet,
which amounts to setting $t(\mathbf{r})=1$ throughout.
Since actin polarization is now no longer localised at the leading edge, we 
instead localise the treadmilling parameter $w$ to be large near the substrate 
to produce a protrusion as in the quasi-2D systems.
While these approximations are useful to make progress in 3D, as the
numerics is more demanding there, we note that the protrusion
transition found in Fig.~1 is still preserved, both
in 3D and in 2D when similar approximations are made (Supplementary Fig.~1).

The passive stress consists of a viscous term, written as  $\sigma_{\alpha\beta}^{viscous}=\eta(\partial_{\alpha}v_{\beta}+\partial_{\beta}v_{\alpha})$, where $\eta$ is the shear viscosity (the greek
indices denote cartesian components), summed with an elastic stress $\underline{{\underline{\sigma}}}^{elastic}$, due to elastic deformations in the liquid crystal phase, and a surface tension term 
$\underline{{\underline{\sigma}}}^{interface}$.  The elastic stress is given by
\begin{eqnarray}
\sigma_{\alpha\beta}^{elastic}=\frac{1}{2}(P_{\alpha}h_{\beta}-P_{\beta}h_{\alpha})-\frac{\xi}{2}(P_{\alpha}h_{\beta}+P_{\beta}h_{\alpha})
                           -\kappa\partial_{\alpha}P_{\gamma}\partial_{\beta}P_{\gamma}\label{eq:elastic-stress},
\end{eqnarray}
similar to that used in nematic liquid crystal dynamics. Here the constant $\xi$ depends on the geometry of the active particles: $\xi>0$ describes rod-like particles, $\xi<0$ oblate ones. We have set it positive, as in~\cite{elsen}.
In addition $\xi$ controls whether particles are flow aligning in shear flow ($|\xi|>1$), hence creating a stable response, or flow tumbling ($|\xi|<1$), which gives an unstable
response. Here we assume $\xi>1$. Note that the elastic stress also depends on the molecular field $h$, defined as $\delta F/\delta\mathbf{P}$.
The surface tension term takes into account interfacial contribution between the passive and the active phase, and is given by
\begin{equation}
\sigma_{\alpha\beta}^{interface}=\left( f-\phi\frac{\delta F}{\delta\phi} \right)\delta_{\alpha\beta} - \frac{\partial f}{\partial\left(\partial_{\beta}\phi\right)} \partial_{\alpha}\phi.
\end{equation}
This term is borrowed from binary fluids~\cite{deGroot-Mazur}, with $f$ being the free energy density. 

The dynamics of the concentration of active material $\phi(\mathbf{r},t)$ is governed by a convective-diffusion equation
\begin{equation}\label{conc_eqn}
\frac{\partial \phi}{\partial t}+\nabla\cdot\left(\phi(\mathbf{v}+w_0\mathbf{P})\right)=\nabla\left( M\nabla\frac{\delta F}{\delta \phi}\right)
\end{equation}
where $M$ is a thermodynamic mobility parameter, related to the diffusion constant $D\simeq Ma$ (see Eq.(\ref{fe})).
$\delta F/\delta \phi$ is the chemical potential of the system and $w_0$ is the self-advection parameter. 
This $w_0$ can be interpreted as the speed, relative to the bulk fluid velocity $\mathbf{v}(\mathbf{r},t)$, at which each filament is self propelled along its own tangent~\cite{elsen}.
Thus $w_0$ is also proportional to the rate of actin treadmilling.
In our 3D minimal model, the treadmilling rate is a spatially varying function which is large close to the substrate:
$w_0\rightarrow w(\mathbf{r})=w_0\exp(-z/\lambda)$ where $\lambda$ is the lengthscale for protrusion thickness.

The last ingredient to complete the description of the physics of the system is an evolution equation for the polarization field $\mathbf{P}(\mathbf{r},t)$. This is
given by
\begin{eqnarray}\label{P_eqn}
\frac{\partial\mathbf{P}}{\partial t}+\left((\mathbf{v}+w_0\mathbf{P})\cdot\nabla\right)\mathbf{P}=-\underline{\underline{\Omega}}\cdot\mathbf{P}+\xi\underline{\underline{D}}\cdot\mathbf{P}
-\frac{1}{\Gamma}\frac{\delta F}{\delta\mathbf{P}}\label{Pdot},
\end{eqnarray}
where $\Gamma$ is the rotational viscosity and $\underline{\underline{D}}=(\underline{\underline{W}}+\underline{\underline{W}}^T)/2$ and 
$\underline{\underline\Omega}=(\underline{\underline{W}}-\underline{\underline{W}}^T)/2$ 
are the symmetric and the anti-symmetric part of the velocity gradient tensor $W_{\alpha\beta}=\partial_{\alpha}v_{\beta}$.
Note that this equation is the one usually employed in polar liquid crystal theory~\cite{Marchetti}.

To solve these equations we use a 3D hybrid lattice Boltzmann algorithm already successfully tested in other systems, such as binary fluids~\cite{pre}, 
liquid crystals~\cite{softmatter} and active matter~\cite{elsen}. It consists in solving Eq.(\ref{conc_eqn}) and Eq.(\ref{P_eqn}) via a finite difference
predictor-corrector algorithm while the integration of the Navier-Stokes equation is taken care of by a standard Lattice Boltzmann approach.
This numerical approach is necessary due to the couplings between velocity field and polarization field in Eq.(\ref{P_eqn}) and velocity field
and concentration in Eq.(\ref{conc_eqn}).
 
\section*{Acknowledgements}
 Work funded in part by EPSRC Grant EP/J007404/1. We thank R. J. Hawkins for very helpful discussions, and K. Stratford for help with coding. ET thanks SUPA for a prize studentship. MEC holds a Royal Society Research Professorship.

\section*{Author contributions}
All authors designed and performed the research and analysed the data.

\section*{Additional information}

{\bf Competing financial interests:} The authors declare no competing financial interests.

\newpage

\section*{Supplementary Note 1: Boundary conditions and the effect of wall slip}

To simulate a cell crawling on a flat solid surface, we need to set up an 
appropriate treatment of droplet wetting at the substrate. 
To this end, we first initialized our system with a half-spherical droplet on 
the plane, with concentration $\phi=\phi_{eq}$ inside and $\phi=0$ outside. 
The polarization field inside the droplet is initially uniform and points 
along the $y$-direction, while it is zero outside the droplet. 
The droplet is therefore initially polarised; the system
then evolves via the equations of motion described in the previous section.

At the substrate the polarization ${\bf P}$ has a parallel and strong 
anchoring, i.e. $P_{z}(z=0)=0$ and 
$\frac{\partial P_{x}}{\partial z}\bigg|_{z=0}=\frac{\partial P_{y}}{\partial z}\bigg|_{z=0}=0$, while the anchoring of the polarization at the surface of the 
droplet is provided by the term $\beta{\bf P}\cdot \nabla\phi$ in the free 
energy (see the main text). With $\beta>0$, the polarization points outwards and
is perpendicular to the surface of the droplet, while 
with $\beta<0$ it points inwards (remaining perpendicular to the
surface). For the actomyosin concentration $\phi$ at the wall
we adopt neutral wetting which means that ${\bf a}\cdot\nabla\phi |_{z=0}=0$ and ${\bf a}\cdot\nabla(\nabla^2\phi) |_{z=0}=0$, where ${\bf a}$ is a 
unit vector normal to the substrate. 
These conditions together ensure ${\bf a}\cdot\nabla\mu |_{z=0}=0$, so that the 
concentration gradient is parallel to the wall and there is no flux across it.

Importantly, focal adhesions, which prevent the filaments from slipping, are taken into account by using non-slip boundary conditions at the substrate ($z=0$). 
This means that ${\bf v}(z=0)=0$. 
Focal adhesions play a fundamental role in cells sliding on a surface. Their absence can cause the actin filaments to slip backwards 
as they polymerise against the cell membrane. On the other hand, experimentalists can reduce focal adhesions by coating the glass slide with appropriate
enzymes before putting the cell (such as a keratocyte) on top~\cite{Theriot-Mogilner}. In order to emulate the reduced presence or lack of focal adhesions
at the surface, we have also simulated a system in which there is a mixture of non-slip and full-slip boundary conditions (see e.g. Fig. 1 in the main
text, and Supplementary Fig. 1). This is done by following 
the approach proposed in Ref.~\cite{Wolff} in which a parameter $s$ is introduced to model the ``fraction of the slip'' at the wall. More specifically,
a partial slip condition is defined to be
\begin{eqnarray}
v_z(z=0)&=&0\nonumber\\
\frac{v_i(z=0)}{v_i^{\prime}(z=0)}&=&3\eta\frac{s}{1-s},
\end{eqnarray}
where $i=x,y$ and $s$ is the slip parameter ($s\rightarrow 0$ corresponds to non-slip condition while $s\rightarrow 1$ corresponds to full-slip
condition). The ``density'' of focal adhesion between the cell and the 
substrate might be quantified via the term $1/s$ -- if there are fewer focal
adhesions the filaments are more likely to slip. 
Fig.~1C in the main text shows a plot of the crawling speed as a function of 
the slip parameter for a 2D simulation. 
Since the slip velocity at the surface will be in the opposite direction to 
that of the actin polymerization, an increase in the amount of slip causes 
a decrease in the crawling speed. This is consistent with the fact
that the actin polymerization becomes less effective at cell protrusion
with fewer focal adhesions~\cite{Theriot-Mogilner,bray}. 

\section*{Supplementary Note 2: Flow field in the frame of reference of the cell}

The flow fields in Fig.~4 of the main text are measured in the reference frame of the laboratory. Within the frame of reference
of the cell, the flow is mainly backwards (Supplementary Fig.~2), 
in agreement with the simulations of Ref.~\cite{mogilnerflow} 
and with the experimental results in Ref.~\cite{keren} for the case of a cell treated with blebbistatin. Contractility 
experimentally was found to be able to reverse the sense of this flow.
In qualitative agreeement with this finding, we observe that contractility leads to a larger flow
(see the intracellular flow in the 2D simulations in Supplementary Fig.~3, where simulations with wall slip and with $\zeta=0$ or $\zeta\ne 0$
are directly compared). 

Finally, it is possible to make semiquantitative contact with experiments measuring retrograde actin flow by tracking, instead of
${\mathbf u}$, ${\mathbf u} +w{\mathbf P}$. In reality, ${\mathbf P}$ is non-zero only close to the surface, as in our
2D simulations in Fig. 1 of the main text (see also Supplementary Movies 1 and 2). Supplementary Fig.~3 shows the
intracellular flow field for 2D simulations, in the frame of reference of the cell, which equals ${\mathbf u -{\mathbf V}_{\rm cell}}$ (${\mathbf V}_{\rm cell}\simeq w{\mathbf P}$
is the cell velocity). 
In the same frame of reference, the F-actin flow would be
equal to ${\mathbf u} +w{\mathbf P}-{\mathbf V}_{\rm cell}\simeq {\mathbf u}$ (${\mathbf V}_{\rm cell}\simeq w{\mathbf P}$
is the cell velocity). Close to the surface ($z\sim 0$), where the experiments typically track retrograde flow,
the F-actin flow velocity is therefore approximately ${\mathbf u}(z=0)$, hence there is no retrograde flow unless there is
wall slip (as in the case of Supplementary Fig.~3).
This should be expected, because strong focal adhesion leads to no slip,
strong friction, hence no retrograde actin flow. 

\section*{Supplementary Note 3: Oscillatory dynamics and medium-dependent viscosity}

Our minimal physical model supports the formation of 
several 3D cell morphologies, some of them remarkably close to their 
experimental counterparts (see Fig. 2 in the main text). 
We recall here that in our 3D simulations, unlike in typical 2D
simulations (e.g. in Fig. 1 of the main text), we have simplified the 
actomyosin contractility to be uniform throughout the droplet ($t({\mathbf r})=1$). 
However, in our 3D simulations actin polarization is 
confined to a thin layer close to the substrate (unlike the 2D case 
in Fig. 1, where actin polarization and treadmilling are only effective 
at the leading edge of the cell). 
This is achieved by assuming a spatially varying treadmilling whose 
functional form is $w(z)=w_0\exp(-z/\lambda)$, where $z$ is the Cartesian 
coordinate perpendicular to the substrate and $\lambda$ is related to the 
thickness of the leading edge. 2D simulations using this prescription
lead to a similar transition as observed in Fig. 1 of the main text
(see Supplementary Fig.~1.)

While the structures shown in Fig.~2 and Fig.~3 
are the most common ones, as noted in the main text, these are not
the only ones supported by our model.
In particular, for suitable choices of parameters the steady state is no longer
a stable crawling morphology as described in the main text, but can also become
oscillatory. 

In Supplementary Fig.~4 ($a$-$h$) we show the time evolution of one such oscillatory motion, where the active droplet glides on a flat surface while 
its leading edge continuously switches between two different shapes. 
 A typical crawling mechanism is initially observed ($a$) while, later on, the leading edge of the droplet undergoes a deformation characterized by a depression ($b$) that splits the lamellipodium into two ``feet'' ($c$ and $d$). These
then separate and later on gradually merge ($e$); the droplet then crawls again
with a single lamellipodium and eventually the cycle repeats ($f$, $g$, $h$).
This dynamics is qualitatively reminiscent of the oscillatory crawling motion
of keratocytes on substrates with high friction~\cite{Theriot-Mogilner}.
Similar sequences of shape changes are also observed when cells develop a 
phagocytic cup before engulfing foreign bodies such as solid particles~\cite{Dembo}. Our results therefore show that similar shape changes can be
due to cell elasticity, actin treadmilling and motor activity alone. 

Finally, we come back to the fact that our simulations in the simplest version 
address a droplet moving within a fluid of its same viscosity.
In reality, the two fluids will not have the same viscosity (the cell usually
moves within a thin film of an aqueous medium), and the membrane viscosity
will be significantly larger than that of the cytosol.
To simulate this, we consider a case in which the viscosity is intermediate in the 
cell (the cytoplasm viscosity is $\sim$ 10 cP~\cite{lubyphelps}), high within the
membrane, and low outside. This more complex situation leads again to our
realistic cellular morphologies, an example with a lamellipodium-like 
protrusion is shown in Supplementary Fig.~5.

\section*{Supplementary Note 4: Mapping to physical units}

In this section we briefly review how to map simulation to physical units in
our work (see also Ref.~\cite{elsen} for more details); this mapping
is recapped in Supplementary Table 1.
Parameters have been chosen to be of the same order of magnitude as 
the ones quoted in~\cite{Theriot-Mogilner,mogilnerflow}.
In our case, realistic parameter ranges can be 
achieved by fixing the length-scale, time-scale and force-scale as: 
$L=0.3\,\mu \mathrm{m}$, $T=10\, \mathrm{ms}$, and $F=1000\, \mathrm{nN}$ 
respectively (in simulation units these scales are all equal to one exactly). 
This correspondence then determines the mapping to physical units
of all quantities.

Note that, in line with standard practice in Lattice Boltzmann simulations, 
the fluid mass density $\rho$ is much larger than the actual mass density of a 
real solvent (water)~\cite{codef}. 
This is acceptable so long as inertial effects remain sufficiently small,
and this procedure speeds up the computations by several orders of magnitude. 
In this case the choice of the force/density scale above keeps the Reynolds
number below $0.1$ which is small enough for inertial effects to be negligible.

\newpage

{\bf Figure 1:} 
{\bf Quasi-2D crawling dynamics.} (a) shows a quasi-2D actomyosin droplet crawling along the treadmilling direction (to the right in the picture). 
The black arrows here represent actin polarization which points in the same direction as actin treadmilling.
Additionally, actin polarization is localised at the leading edge of the droplet to model a crawling eukaryotic cell 
(also see Supplementary Movies 1 and 2).
If the treadmilling rate $w_0$ is large enough, a thin protrusion forms, whose steady-state size (area per unit transverse length) is plotted versus $w_0$ in (b) 
(red/plus curve), for the case of no-slip boundary condition between the droplet and the substrate. 
From (b), we can identify the critical treadmilling rate $w_c$ above which a thin layer of protrusion is observed in a crawling cell.
This transition becomes sharper as the amount of slip at the substrate increases (green/cross curve),
and discontinuous/first order for higher values of $s$ (blue/asterisk and purple/square curves). The 
hysteresis loop corroborates its discontinuous character.
(c) shows a plot of the crawling speed as a function of the slip parameter $s$ (while the treadmilling rate is fixed at $w_0=0.002$). 
This controls the discontinuity in fluid velocity at the solid wall, and is defined such that $s=0$ corresponds to no-slip while $s=1$ corresponds to full-slip. 
An increase of $s$ causes a decrease of the velocity of the cell. 
This is because the slip velocity at the surface is in the opposite direction to that of the actin polymerization.
Again, we observe a discontinuity at $s=s_c$ marking a dynamical phase transition from a round shaped crawling cell ($s>s_c$) to a protruded shape ($s<s_c$).

{{\bf Figure 2:} 
{\bf Spreading and crawling cell morphologies.}
In this figure spreading and crawling cell morphologies observed in past experiments (left column) and obtained in the current 3D simulations (middle and right column, giving respectively 
top and side view) are shown. 
Frame a-c shows a static cell with a fried egg shape; the left column panel was originally published in Ref.~\cite{Yam2007}. Other frames show moving cells with protrusions resembling 
respectively, a lamellipodium (d-f; the left column panel was originally
published in Ref.~\cite{Barnhart2011}), a phagocytotic cup (g-i; left column 
panel taken from Ref.~\cite{JCS1} with permission), and
a pseudopod (j-l;  left column panel taken from Ref.~\cite{JCS2} with 
permission). Parameters for the 3D modelling 
(in simulation units; see Supplementary Table 1) 
are: (a-c) $w_0=0.0035$; $\zeta=-0.001$; $\beta=0.02$; (d-f)
 $w_0=0.04$; $\zeta=-0.0015$; $\beta=0.001$; (g-i) $w_0=0.04$; $\zeta=0$; $\beta=0.001$; (j-l)
 $w_0=0.04$; $\zeta=-0.00015$; $\beta=0$.}

{\bf Figure 3:}
{\bf Phase diagram.} 3D Steady state shapes for (a) $\beta=0$, $\zeta=-0.00015$, $w_0=0.04$; (b) $\beta=0.001$, $\zeta=-0.001$, $w_0=0.04$; (c) $\beta=0.001$, $\zeta=-0.0015$, $w_0=0.04$;
(d) $\beta=0.02$, $\zeta=-0.001$, $w_0=0.04$. Red arrows indicate the direction of the polarization in the cytoplasmic projections (psedudopod (a), phagocytinc cup (b), lamellipodium (c)
and fried egg crown (d)). Notice that $w_0$ is constant; while $\zeta$ varies 
(both $\zeta$ and $\beta$ affect droplet elasticity, increasing $\zeta$ for
small $\beta$ leads to further instabilities such as droplet breakup 
which are not relevant for the current work). 

{\bf Figure 4:}
{\bf Polarization and velocity fields.} This figure shows the section of the polarization field inside the droplet very close to the substrate (left column) and the velocity field (right column) for (a)-(b) fried egg, (c)-(d) lamellipodium, (e)-(f) phagocytic cup, (g)-(h) pseudopodium, respectively.

{\bf Figure 5:}
{\bf Flow field of a contractile active droplet.} The figure shows the flow field (in the laboratory frame of reference) for a contractility-driven motile active droplet (left from Ref.~\cite{elsen};
for simplicity in the bulk), and for a cell moving in 3D inside matrigel (from Ref.~\cite{poincloux}, with permission).

{\bf Figure 6:} 
{\bf 3D dynamics in absence of treadmilling.} Snapshots corresponding to the motion of a contractile cell 
on a substrate, in the absence of treadmilling (parameters were
$w=0$, $\zeta=-0.005$, $\beta=0$). The active droplet first elongates perpendicular to the wall
due to the active stress; the polarization then splays in 3D and
this leads to internal flow and a motile concave shape, resembling half of a free-swimming 3D droplet as investigated in~\cite{elsen}.

\newpage

\begin{table}[!h]
\caption{{\bf Supplementary Table 1:} Typical values of the physical quantities used in the simulations.
This choice of parameters are made to be consistent to other physical estimates in~\cite{mogilnerflow,Theriot-Mogilner}.}
\label{table}
\vskip 0.3cm

\begin{tabular}{p{5cm}|c|c}
Model variables and parameters & Simulation units & Physical units  \\
\hline 
Effective shear viscosity, $\eta$ 			 & $5/3$ 		 & $1.86\, \mathrm{kPa}\mathrm{s}$ \\
Effective elastic constant, $\kappa$ 			 & $0.04$ 		 & $40\, \mathrm{nN}$ \\
Shape factor, $\xi$ 					 & $1.1$		 & dimensionless \\
Self-advection/polymerisation speed, $w$	 	 & $0.02$		 & $0.6\,\mu \mathrm{m}\mathrm{s}^{-1}$ \\
Effective diffusion constant, $D=Ma$			 & $0.004$ 		 & $0.036\,\mu \mathrm{m}^{2}\mathrm{s}^{-1}$ \\ 
Rotational viscosity, $\Gamma$ 				 & $1$ 			 & $1.1\, \mathrm{kPa}\mathrm{s}$ \\
Activity, $\zed$ 					 & $0-0.001$ 		 & $(0-10)\, \mathrm{kPa}$ \\
\end{tabular}
\end{table}

\newpage
\begin{figure}[!h]
\begin{center}
\includegraphics[width=15.cm]{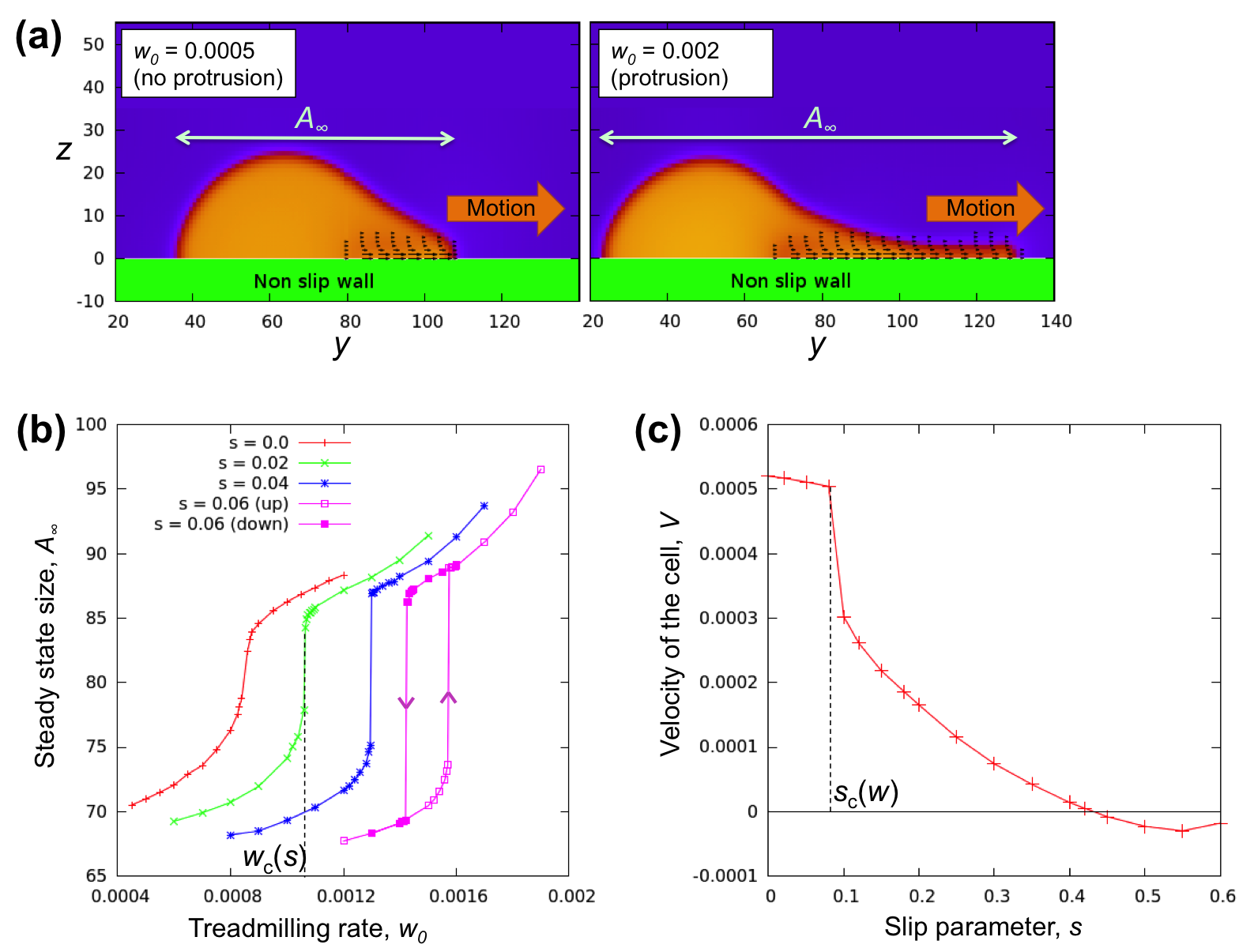}
\end{center}
\caption{Figure 1}
\end{figure}

\newpage

\begin{figure}[!h]
\includegraphics[width=15.cm]{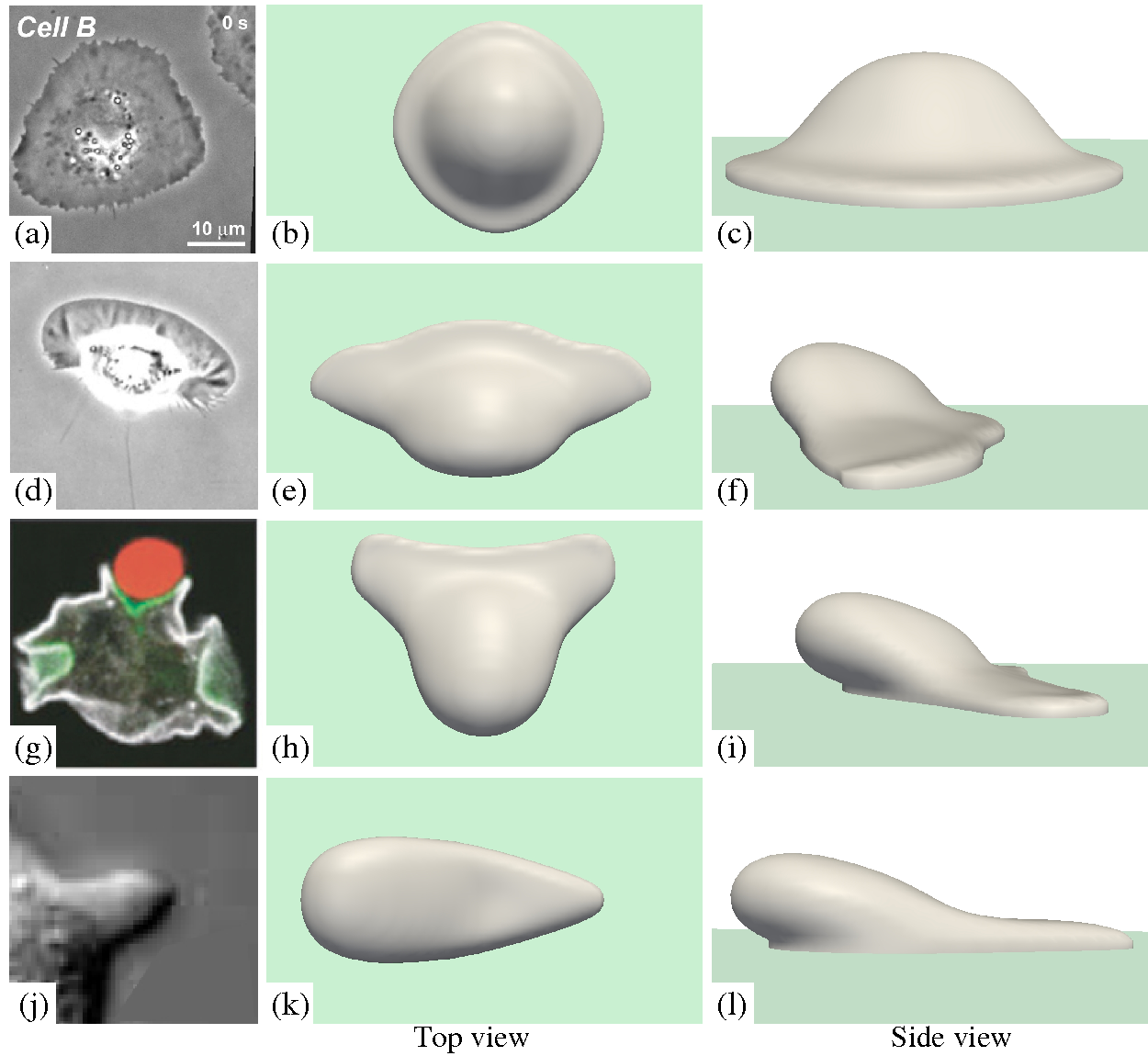}
\caption{Figure 2}
\end{figure}

\newpage

\begin{figure*}[!h]
\centerline{\includegraphics[width=1.4\textwidth]{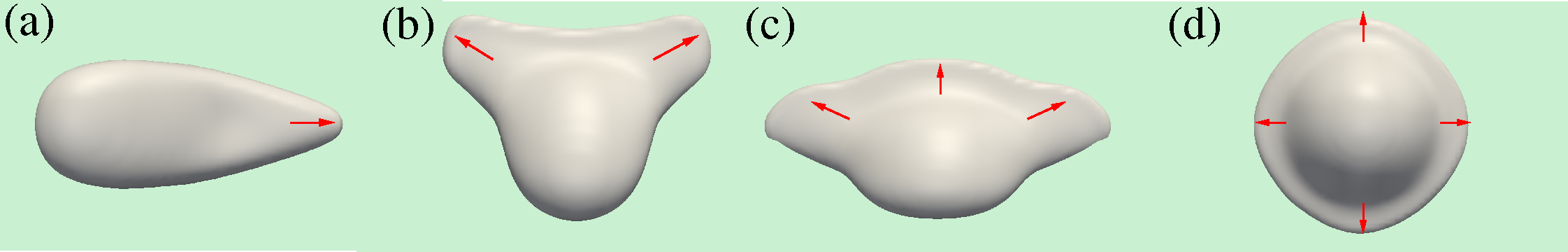}}
\caption{Figure 3}
\label{}
\end{figure*}


\begin{figure*}[!h]
\centerline{\includegraphics[width=1.05\textwidth]{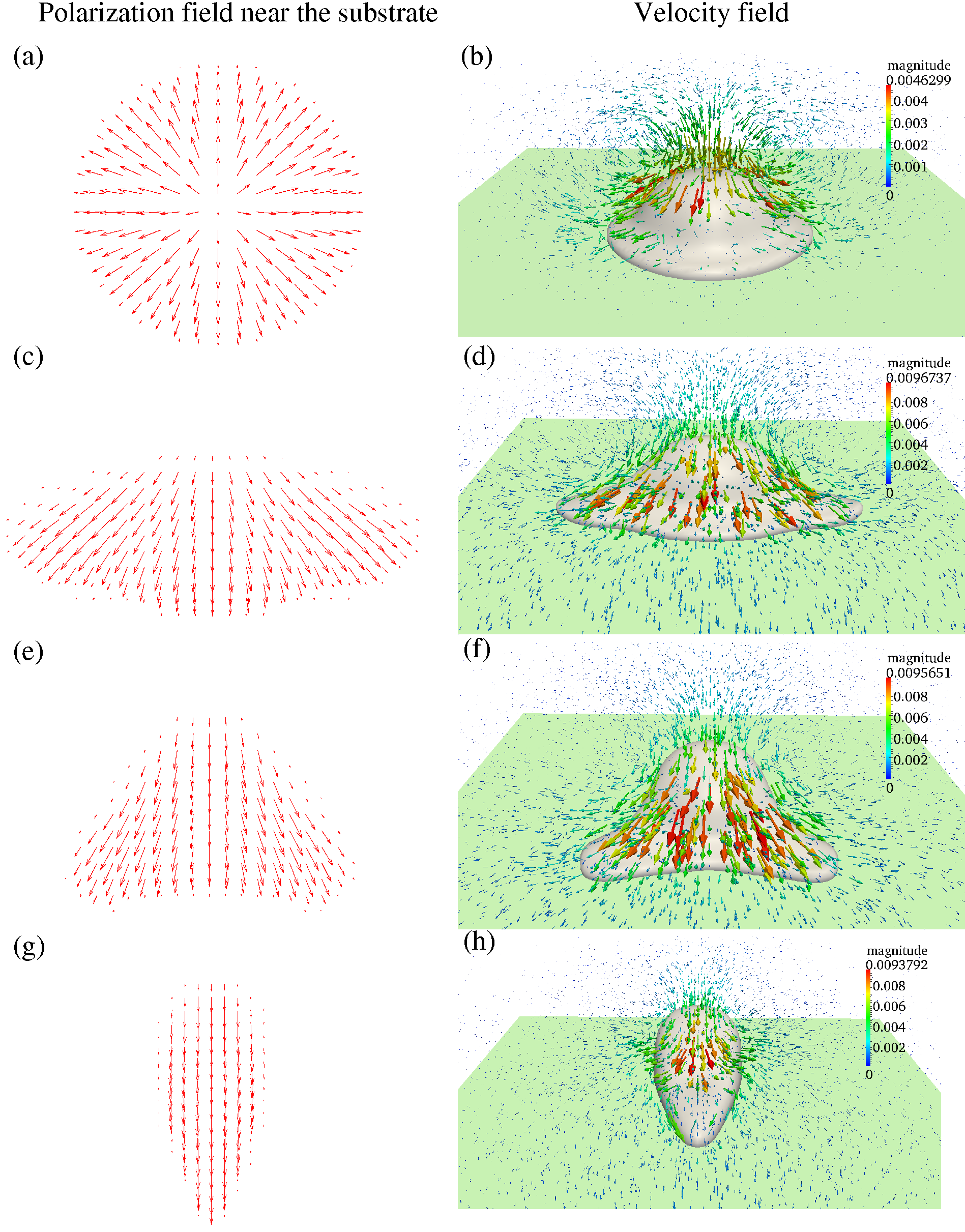}}
\caption{Figure 4}
\label{supp_figure5}
\end{figure*}

\newpage

\begin{figure*}[!h]
\centerline{\includegraphics[width=1.3\textwidth]{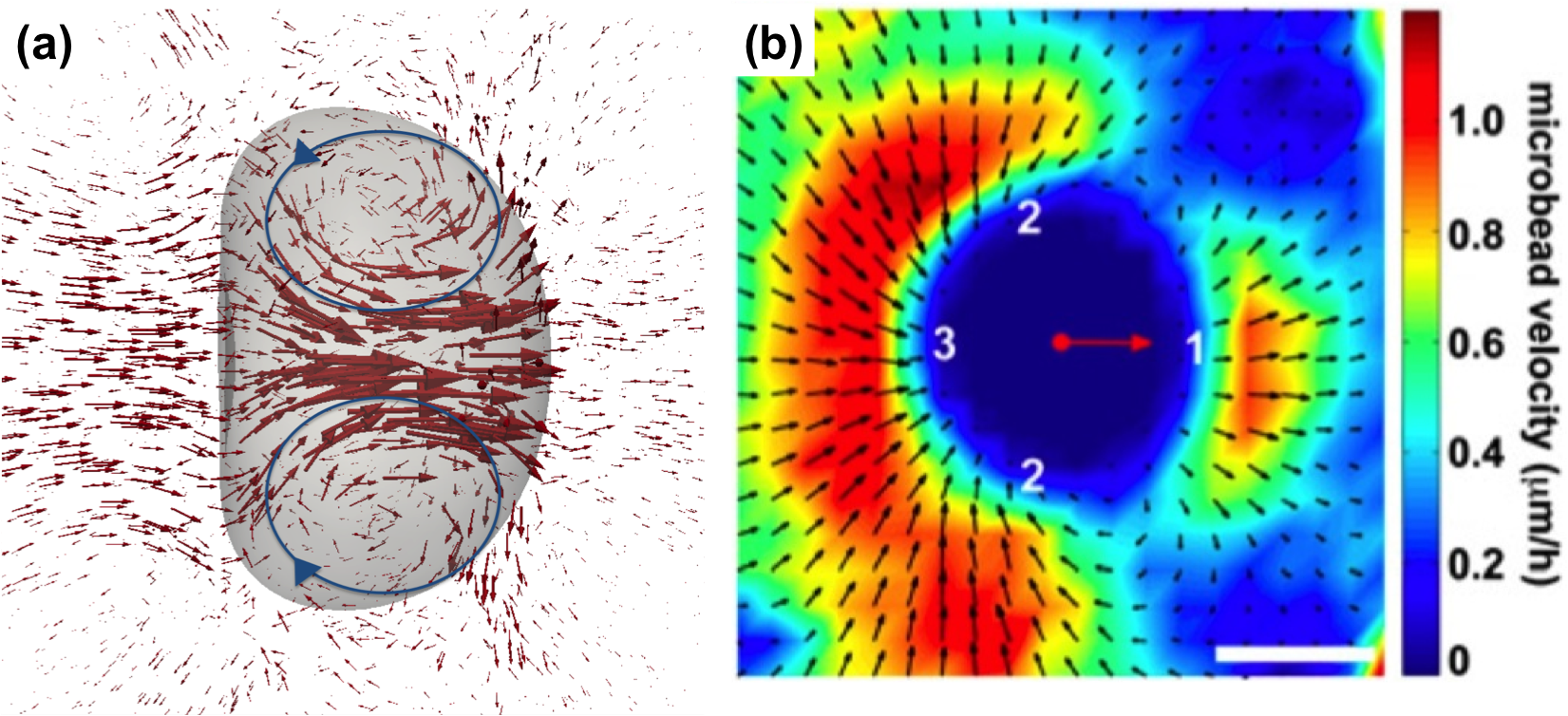}}
\caption{Figure 5}
\label{supp_figure6}
\end{figure*}

\newpage

\begin{figure}[!h]
\begin{center}
\includegraphics[width=15.cm]{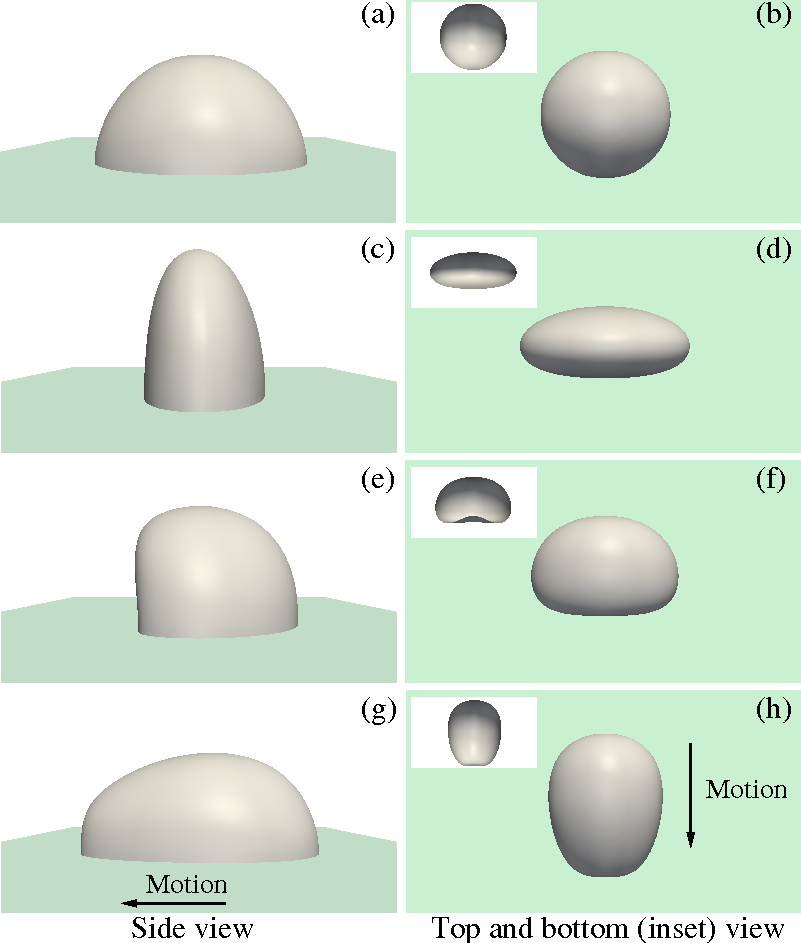}
\end{center}
\caption{Figure 6}
\end{figure}

\newpage

\begin{figure*}[!h]
\centerline{\includegraphics[width=1.\textwidth]{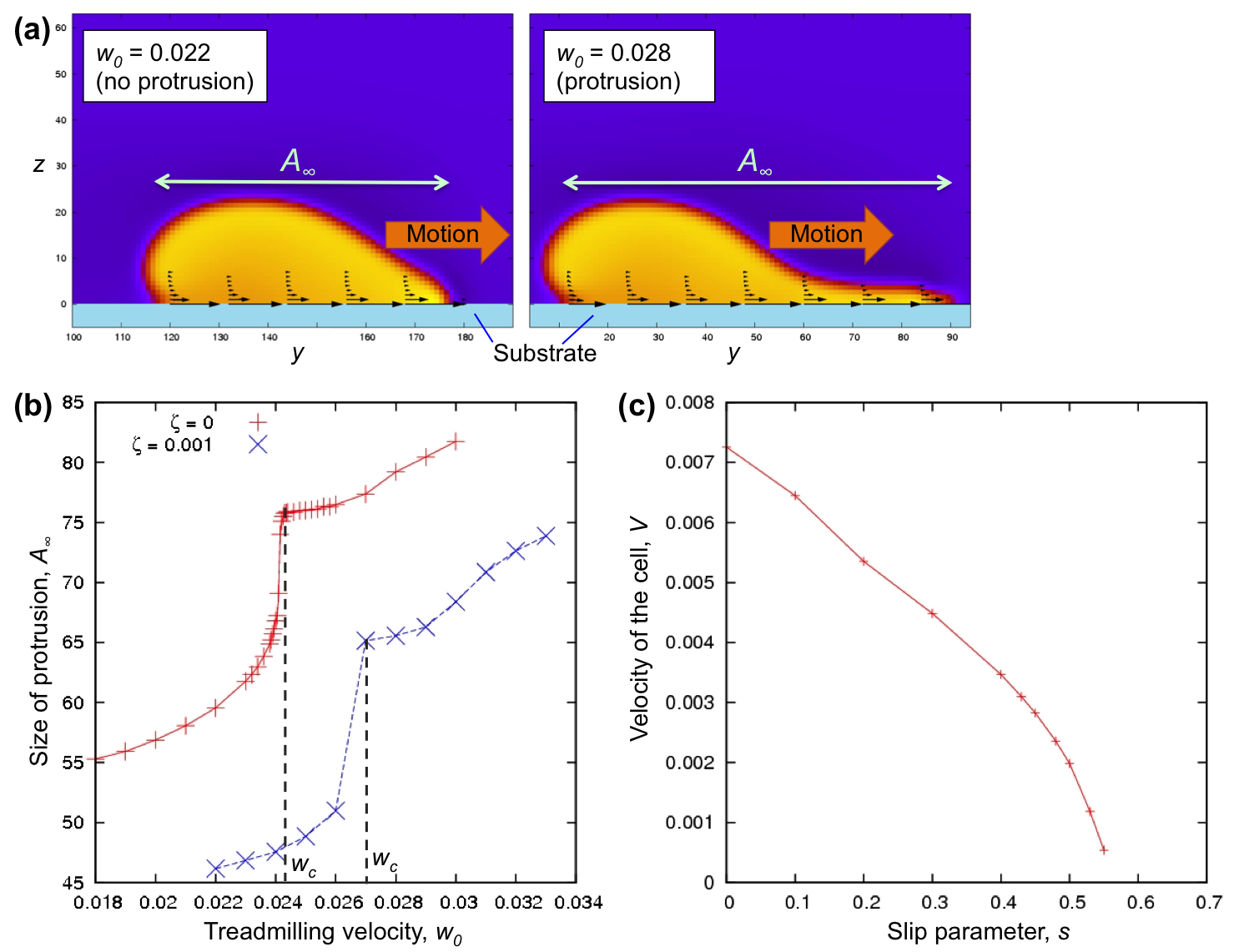}}
\caption*{
{Supplementary Figure 1:} {\bf Simulations with $t({\mathbf r})\equiv 1$.}
(a) Snapshots of a quasi-2D actomyosin droplet crawling along
the treadmilling direction (to the right in the picture). 
There is a transition between a relatively underformed state and
a structure with a thin protrusion, for large $w_0$, as in Fig. 1
in the main manuscript. Unlike that case, we chose $t({\mathbf r})\equiv 1$
(see Methods in the main text); furthermore the self-advection,
or treadmilling, is no longer localised close to the cell leading edge, and is
now equal to $w=w_0\exp(-z/\lambda)$ where $\lambda$ is the 
protrusion thickness.
The steady-state size (area per unit transverse length) is plotted versus $w_0$ in (b). (c) shows a plot of the crawling speed as a function of the slip parameter $s$ 
(see Supplementary Note 1).
}
\label{supplfig1}
\end{figure*}

\newpage

\begin{figure*}[!h]
\centerline{\includegraphics[width=1.3\textwidth]{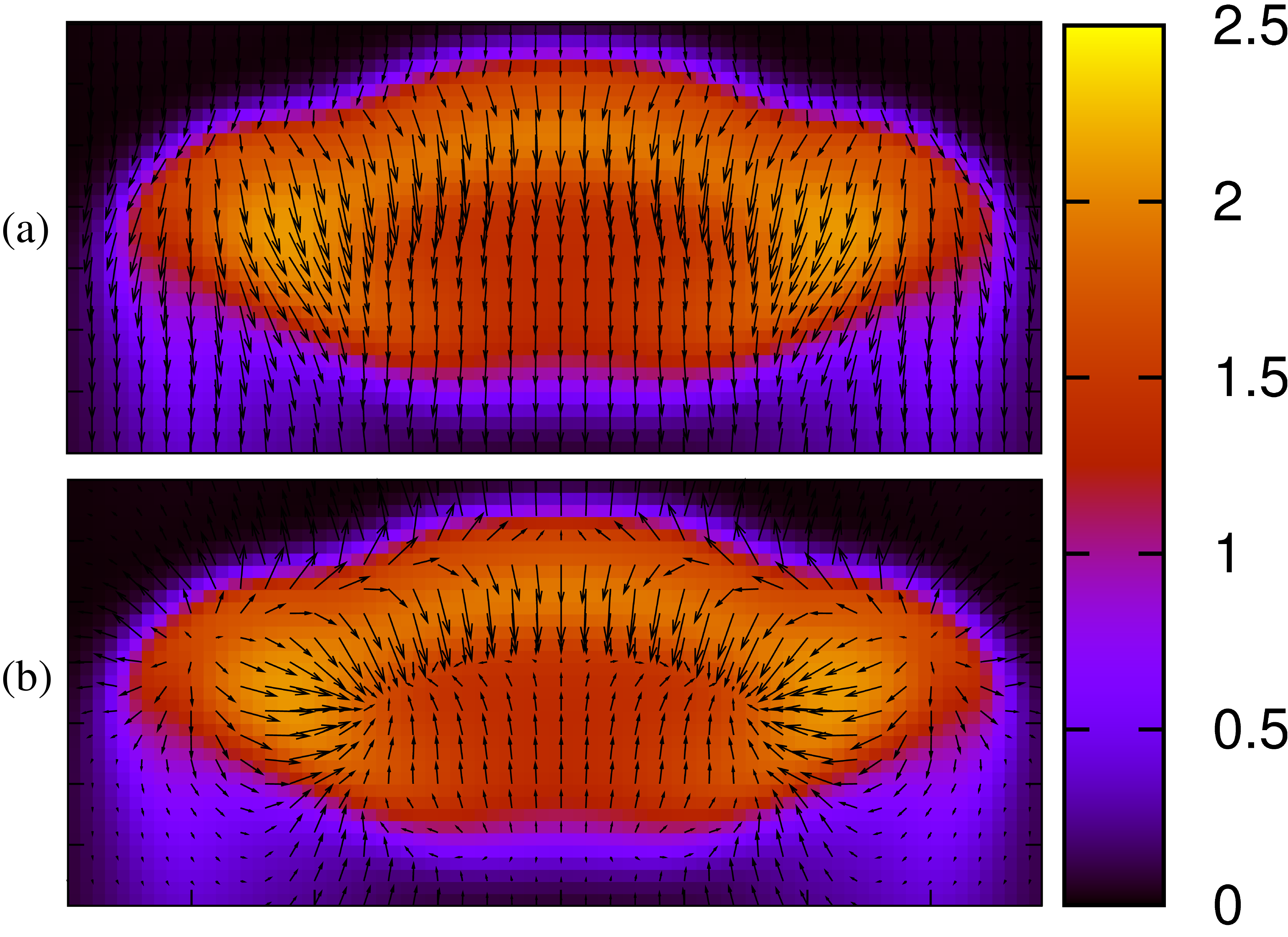}}
\caption{
{Supplementary Figure 2:}
{\bf Flow field in 3D close to the wall.}
In this figure we show a 2D section of the 3D flow field calculated in the 
reference frame of the cell (a) and of the laboratory (b),
in a layer $0.3$ $\mu$m away from the wall, for the 
``lamellipodium'' state in Fig. 2.  This flow field
does {\it not} include the actin flux, hence it should be thought of
as an intracellular solvent flow. 
In the cell frame of reference the flow is almost unidirectional, directed
from the leading edge towards the rear part of the cell,
in agreement with the results presented in Ref.~\cite{keren}. 
}
\label{supplfig2}
\end{figure*}

\newpage

\begin{figure*}[!h]
\centerline{\includegraphics[width=1.3\textwidth]{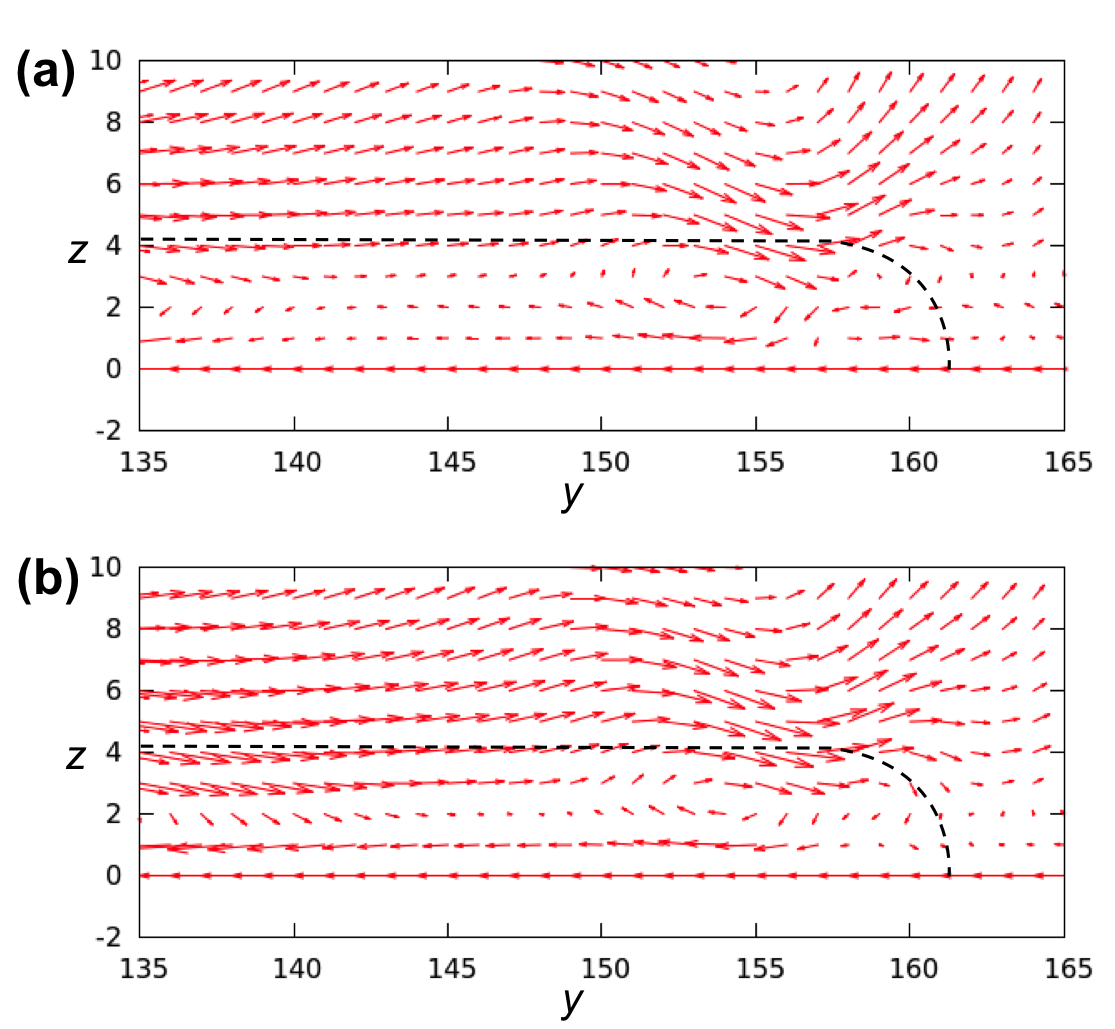}}
\caption{
{Supplementary Figure 3:}
{\bf Intracellular flow with and without contractility.}
Plot of the intracellular flow in the cell frame of reference,  
${\mathbf u}-{\mathbf V}_{\rm cell}$ for a 2D simulation with wall slip, and $\zeta=0$ (top)
or $\zeta=-0.01$ (bottom). 
Contractility leads to a stronger intracellular flow. The retrograde flow
is approximately equal to the velocity field at the substrate (in the lab frame, see text), so it is approximately equal to the slip velocity at the wall.
}
\label{supplfig3}
\end{figure*}

\newpage

\begin{figure*}[!h]
\centerline{\includegraphics[width=1.4\textwidth]{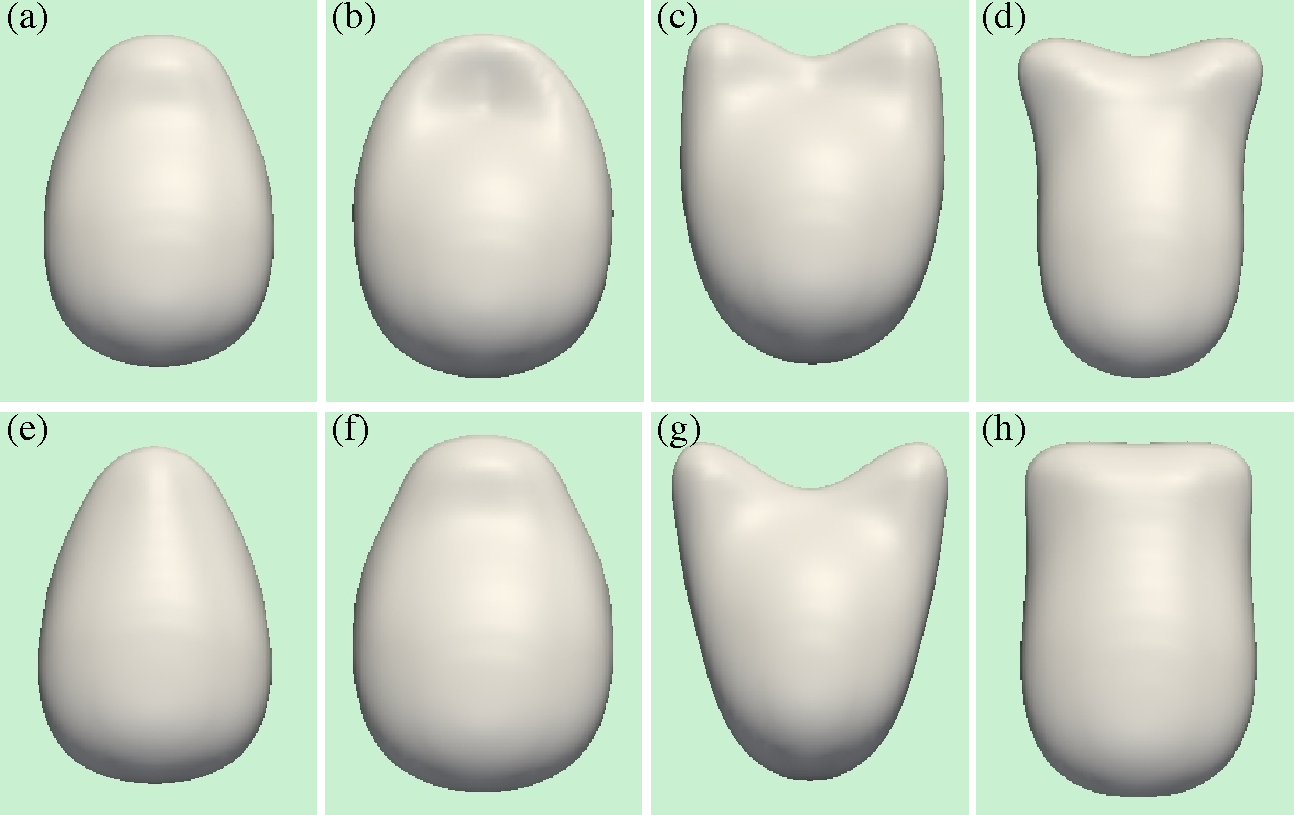}}
\caption{
{Supplementary Figure 4:} {\bf Oscillating active droplets.}
This figure shows the time evolution of
an oscillating active droplet crawling on a flat surface. The movement resembles
the dynamic of a phagocytic cell engulfing a solid particle. The initial crawling seen in ($a$) is characterized
by a small lamellipodium that, later on, elongates and splits into two lamellipodia ($b$, $c$ and $d$). Then they gradually merge ($e$)
until the oscillatory behaviour restarts ($f$, $g$ and $h$).
}
\label{supplfig4}
\end{figure*}

\newpage

\begin{figure*}[!h]
\includegraphics[width=16.cm]{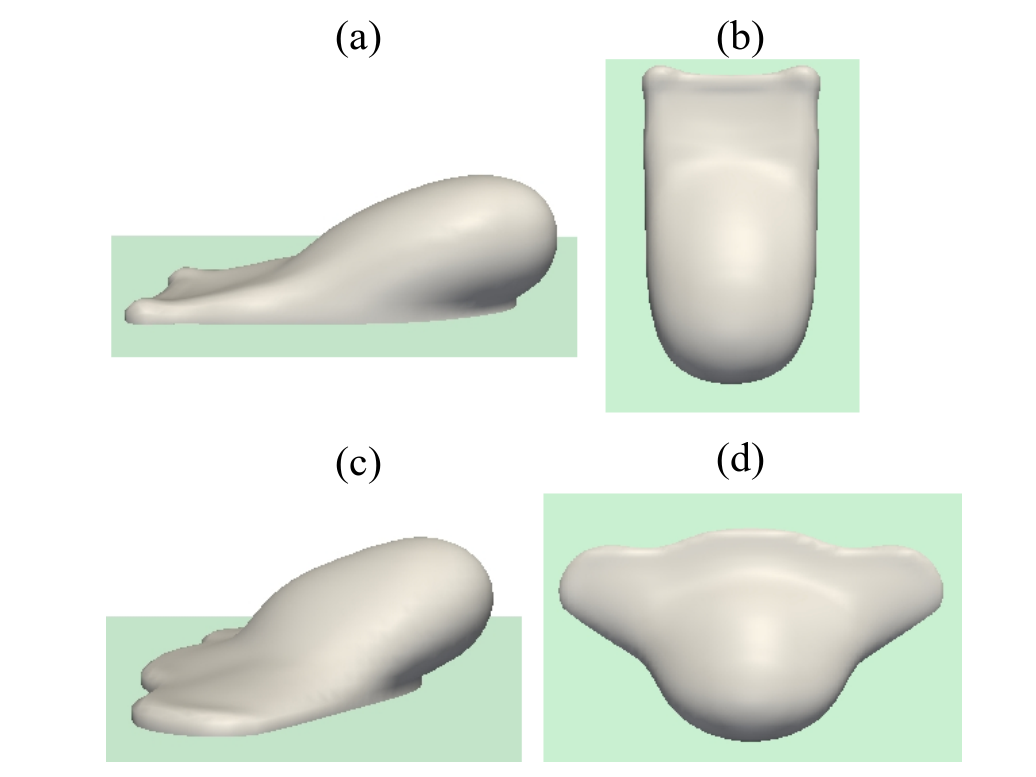}
\caption{
{Supplementary Figure 5:} {\bf Three-viscosity simulations.}
Side view of a cell with a protrusion, in a three-viscosity simulation. Viscosities were (in simulation units):
$\sim 1.67$ (within the cell),  $\sim 6.67$ (within the membrane) and  $\sim 0.53$ (outside the cell). Parameters were: $\zeta\simeq 0$, $\beta=0.0005$, $w_0=0.05$.
}
\label{supplfig5}
\end{figure*}

\end{document}